\documentclass[12pt]{article}

\usepackage{sbc-template}
\usepackage{graphicx,url}
\usepackage{xcolor}

\usepackage[utf8]{inputenc}  

\date{February 2023}

\title{Where are the marathon Girls?: An Analysis of Female Representation in the Brazilian ICPC Programming Marathons}

\author{Crishna Irion\inst{1}, Luiz Claudio Theodoro\inst{1}, \\Flávio de Oliveira Silva\inst{1}, João Henrique de Souza Pereira\inst{1} }

\address{Faculdade de Computação\\ Universidade Federal de Uberlândia(UFU)- MG - Brasil 
  \email{crishna@ufu.br, luiz.theodoro@ufu.br, flavio@ufu.br, joaohs@ufu.br}
}

\begin{document} 

\maketitle

\begin{abstract}
Education motivated the encouragement of female participation in several areas of science and technology. Programming marathons have grown over the years and are events where programmers compete to solve coding challenges. However, despite scientific evidence that there is no intellectual difference between genders, women's participation is relatively low. This work seeks to understand the reason for this adherence, considering the gender issue in Programming Marathons over the last years, in a real context. This work aims to understand the context of female representativeness in which the intellectual aspects do not differ in gender. Still, there is a considerable discrepancy in female belonging.

\end{abstract}

\section{Introduction}

Women, regardless of social class, have historically had their jobs performed with many challenges. Only from the last century, in the 1990s, it was possible to perceive an increase in the incentive for women to enter the labor market, seeking their social and financial emancipation.To \cite{maia_2016} the scenario of female rarity contributes to the continuity of the stereotyped perception of women.

Education was an important instrument in this process, opening the way for achievements in science, exact sciences and technology. However, women have always had to fight many challenges and barriers such as wage differences, physical intimidation, intellectual disqualification and even harassment in order to gain a foothold professionally. 
Despite so many challenges and inequalities, women continue, little by little, to seek their insertion in the labor market.

Participation in scientific events, challenges, Programming Marathons are ways of stimulating the process and building knowledge. The Programming Marathon is an event where programmers compete to solve coding challenges in a short period of time. Although this intellectual contest is growing in popularity, participation by women is relatively low.

The ACM International Collegiate Programming Contest (ICPC) is an annual programming competition between universities around the world. Headquartered at Baylor University and autonomous regions on every continent, the ICPC is organized by the Association for Computing Machinery. In Brazil, the Programming Marathon is promoted by the Brazilian Computing Society and has been held since 1996. The Marathon was created with the aim of classifying regional teams for ICPC world competitions, as part of the South American regional competition.
Women represent less than 15\% of the participants in the Programming Marathons, and most of them are young students. To \cite{maia_2016} the scenario of female rarity contributes to the continuity of the stereotyped perception of women. Although there is still a long way to go in terms of gender equality in Programming Marathons, some progress has been made in recent years regarding initiatives for the inclusion of women in information technology and in technical and higher courses in computing. Some initiatives such as Include Girls, Girls in Tech, CodeGirl, among many others, encourage women and girls to be protagonists in the technology field.

The research considers gender issues in the Programming Marathons, in the first national phase, which is of broad competition, in which teams from schools throughout Brazil, whether private or public, can participate. The objective of this work is to understand female representation in high-performance programming competitions, in which the intellectual aspects do not differ by gender, but there is a huge discrepancy in female belonging.

\section{Related Works} \label{sec:firstpage}

Searches were carried out in the BDTD (Base of National Theses and Dissertations), as well as in Google Scholar, in the events WIT (Women in Technology), the CBIE (Brazilian Congress of Informatics in Education) and the ANPED (National Association of Graduate and Research in Education), with the aim of obtaining studies that analyze the difficulties faced by students in computing courses in the national scenario. The quality criteria used in the selection of works were rigor, credibility and relevance.

\cite{Santos:21} and \cite{maia_2016} analyzed the female scenario of higher IT courses in Brazil, concluding the evident inequality of freshmen and graduates from the years 2014 to 2019.

For \cite{Paganini:20}, in her study on the low participation of women in Hackathon competitions, analyzing points such as motivational aspects and gender problems are key issues to find ways to understand this female scarcity in events, as well as the lack of of empirical evidence studies on the reasons for this.

However, unlike works presented previously, this article presents an analysis of female participation in Programming Marathons, which are high-performance competitions, showing the importance of this research, never before carried out in Brazil, seeking to think of ways to expand this representativeness and female participation.

\section{Methodology}

This research problem is to understand the  women in Programming Marathons participation, looking to understand the representation so far to men along the history of ICPC marathons, of last 15 years .

To meet this article proposed objectives, a bibliographical research was carried out theoretically substantiating the \cite{LAKATOS:2021} study. The search for keywords was: women in IT, women in higher education courses in computing and IT, women in Programming Marathons, gender inequality in computing.

The quantitative analysis of data from IT Courses in Brazil was based on data from 2011 to 2020 provided by \cite{SBC2:23}, as well as the analysis of ICPC data from Programming Marathons (2008 to 2022). For this purpose, the documentary research \cite{Gil:2019} was adopted, which uses materials that have not yet received analytical treatment of the data, and can be treated in order to meet the research objectives. 
The data were made available by the organization of the Programming Marathon, being the official data of the ICPC (2023) \cite{ICPC:2023}, referring to Brazil in the years 2008 to 2022 of the first phase and 2008 to 2021 for the national finals, There are an exception of the 2010 data in the data composition, which meet the research object of this work. The statistical analysis of the data was developed using an algorithm in the Python programming language, due to the large volume of spreadsheets made available by the ICPC, with hundreds of lines, which facilitated the analysis and generation of statistical graphs.

\section{Women Representativeness} 

This research seeks to understand female representation scenario in Programming. Therefore, it was necessary to analyze the scenario of women in computing, their participation in IT courses and the inequality scored in the numbers.

\subsection{Representativeness in Higher Education Courses of Information Technology (IT) in Brazil}
Gender and its categories are represented as a “natural expression of innate sexual differences (for example, women are soft and delicate while men are strong \cite{amaral:17}, showing that they are socially and culturally defined in this way.

Also according to \cite{amaral:17}, the low percentage of women in science was attributed to the natural differences between men and women, but the increased representation of women in some areas shows that the origin of the problem is not linked to the nature of each of the professionals, nor the ability to produce knowledge, but that comes with the time defined from the first steps with the family and then at school. 
Concepts such as the difference between things for boys and things for girls occur from early childhood, in toys and games, worsening over time when the computer becomes part of the boy's context, even if the family has a girl with an interest in technology, according to \cite{amaral:17}.

The role of the girls ends up being that of proximity to the computer only when it is essential for some obligatory activity. And so it happens with the types of toys that each gender receives in childhood: they are “boys toys” animals, vehicles, electronics, assembly games, in contradiction to “girls toys” that get dolls, houses and household items, according to \cite {Nash:17}.

The naturalization of differences between men and women, as well as the intellectual and moral hierarchy between human groups, were projects that had cultural and social reinforcement through practices and techniques of physical anthropometry, such as craniometry and phrenology, since the 18Th century. Darwinian theories of evolution reinforced these issues through ways that emphasized the origin and reinforcement of differences in body parameters metriced by these theories, according to \cite{Sepulveda:22}.

For \cite{Santos:21}, in their study about higher IT courses in Brazil female scenario, analyzing INEP data from 2014 to 2019, shows that the number of technology women students ranged from 13.8\% and 15.2\% studying and graduated, respectively. Gender inequality is evident both in the number of people studying and in the number of graduated.

\begin{figure}[ht!]
\centering
\includegraphics[width=.95\textwidth]{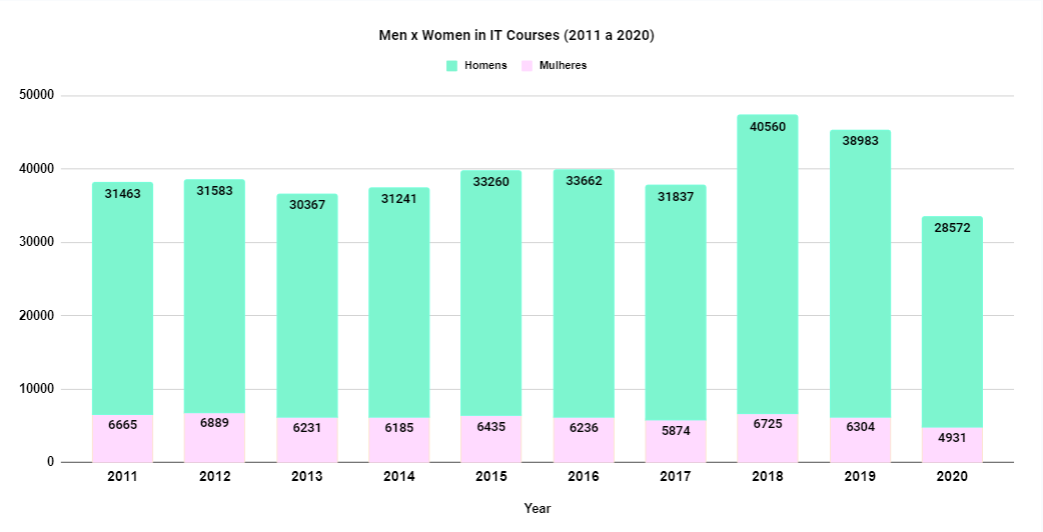}
\caption{Comparison between men and women in IT courses}
\label{fig:exampleFig1}
\end{figure}

In the focus of this research, analyzing data from the years 2011 to 2020 on IT courses, available at SBC, one can see this stability of female representation over the years. However, in addition to there being no growth in the percentage of women, there is still a slight drop in the overall percentage, as can be seen (Figure~\ref{fig:exampleFig1}) in the comparison of men and women \cite{SBC2:23}.

The table \ref{tab:exTable1} shows the data of graduates in higher education courses in Information Technology from 2011 to 2020, which are treated in the graph.

\begin{table}[ht]
\centering
\caption{Grand total of graduates in higher education courses in Information Technology}
\label{tab:exTable1}
\includegraphics[width=.75\textwidth]{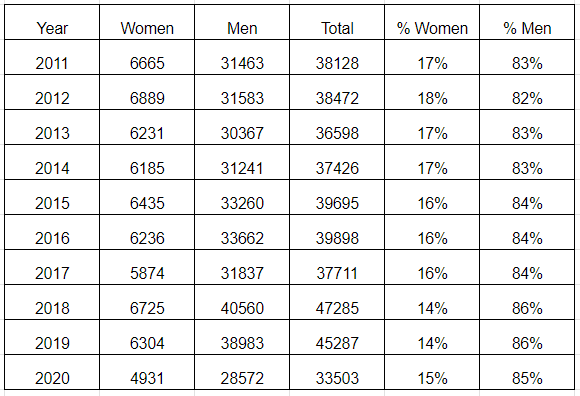}
\end{table}

\cite{amaral:17} suggest, as a perception of their research, that it is essential to investigate the reasons for the lack of motivation among computer science students, even though there are so many professional opportunities currently in the job market.

\subsection{Women Representation in Programming Marathons}
The Programming Marathons are competitions composed of several teams of undergraduate students in Information Technology area with the objective of solving problems and designing computational algorithms.

The Brazilian Computing Society (SBC) Programming Marathon takes place annually, having started in 1996, and is part of the classification stage for the world finals of the International Collegiate Programming Contest (ICPC), organized by Baylor University and partner institutions throughout of the \cite{SBC:23} world.
 
The Brazilian Computer Olympiad (OBI) is an initiative of the Brazilian Computing Society that aims to stimulate interest in Computing and Science, as well as stimulating computational reasoning and programming techniques among high school and elementary school students.

Over the years, female participation in computing courses has been the subject of discussion, as well as in technical courses in the area of technology. Many initiatives have been created and developed to bring girls and women closer to the technology field. 

However, this same initiative has generally not taken place in Brazil for programming competitions. Even with several female highlights in competitions, whether in the OBI (Brazilian Informatics Olympiad for the elementary and middle school years), or in the programming Marathons, few movements have taken place in the sense of empowering and attracting the female public to participate.

An objective that matches the perception of this study is "to identify talents and vocations in Computer Science in order to better instruct and encourage them to pursue careers in the areas of science and technology” \cite{OBI:22}.

A brief and superficial analysis of the OBI 2022 medalists, analyzing only the initiation levels level 1 (sixth and seventh years of Elementary School) and senior level programming (fourth year of Technical Education and students attending the first year of a course for the first time of graduation), there is some evidence regarding the interest, motivation and participation of girls in this type of intellectual competition.

To give you an idea, this year (2022), the OBI had a total of 98 medalists in the Level 1 Initiation Modality, in which 28 girls were in this ranking, with the first girl being placed 4Th in the competition. This corresponds to a little more than 28\% representation among 11 and 12 year old girls. 
In Senior Level Programming Modality, which has young participants aged just over 16, only one girl was a medalist, out of a total of 51 participants, which corresponds to 1.96\% of female participation.

\section{Results and Discussion}\label{sec:figs}

\begin{figure}[ht!]
\centering
\includegraphics[width=.8\textwidth]{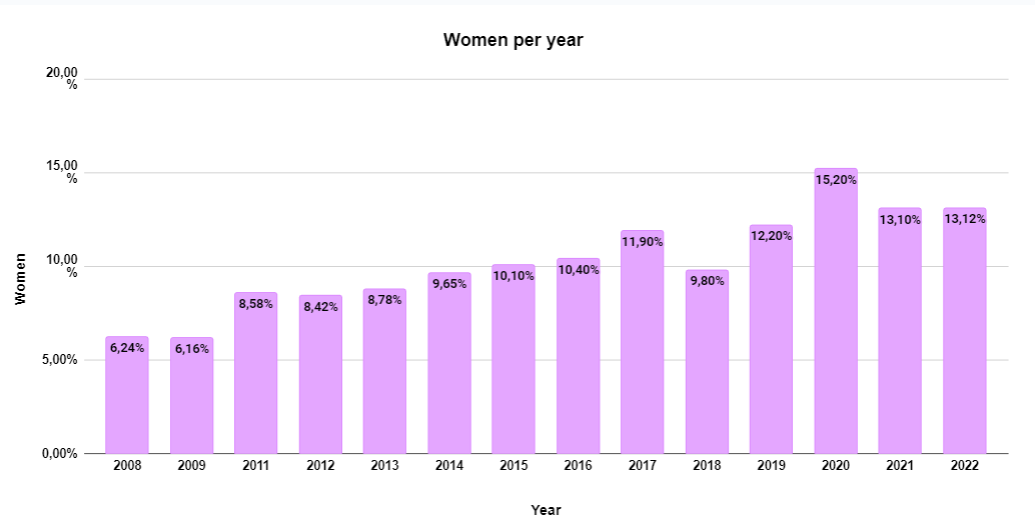}
\caption{Number of Women in the first phase per year}
\label{fig:exampleFig2}
\end{figure}

At first, the analysis of  data is very positive regarding the participation of women in the Programming Marathon. The first phase of the marathon is a democratic competition, all teams from Brazilian schools, private or public, can participate. Students of all computing courses have the opportunity to be there.
The first figure (Figure~\ref{fig:exampleFig2}) shows the absolute number of female participants per year.

At first, it is possible realize an increase about women participation per year, especially in 2020.
In percentage terms, the comparison of female participation, in terms of gender, shows that this evolution has a percentage variation that oscillates between 4.24\%. 

\begin{figure}[ht!]
\centering
\includegraphics[width=.8\textwidth]{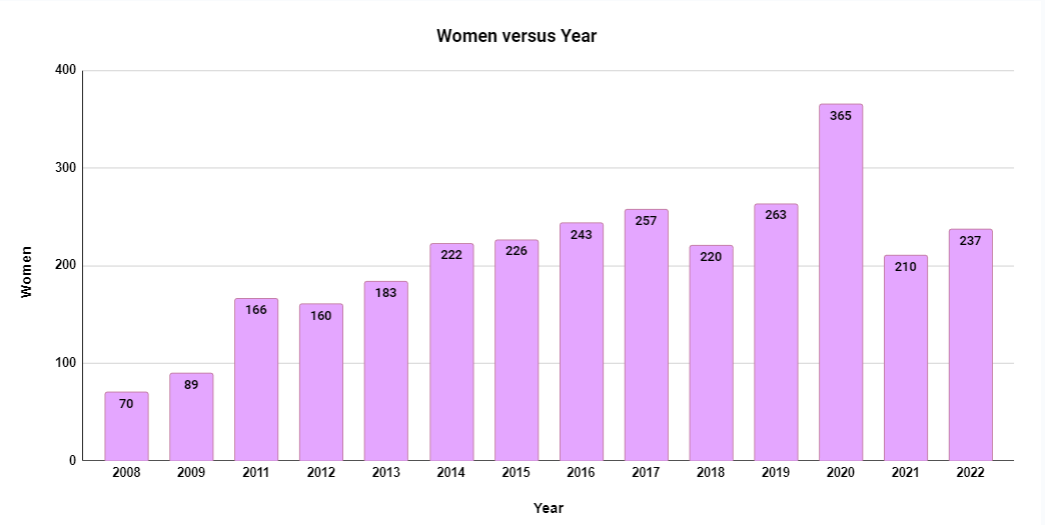}
\caption{Women Percentage in the marathon first phase(2008 to 2022)}
\label{fig:exampleFig3}
\end{figure}

The worst representation was in 2008, and 13.82\%, the highest percentage in 2020, according to (Figure~\ref{fig:exampleFig3}) which shows the percentage of female participants in the first phase per year.

\begin{figure}[ht!]
\centering
\includegraphics[width=.9\textwidth]{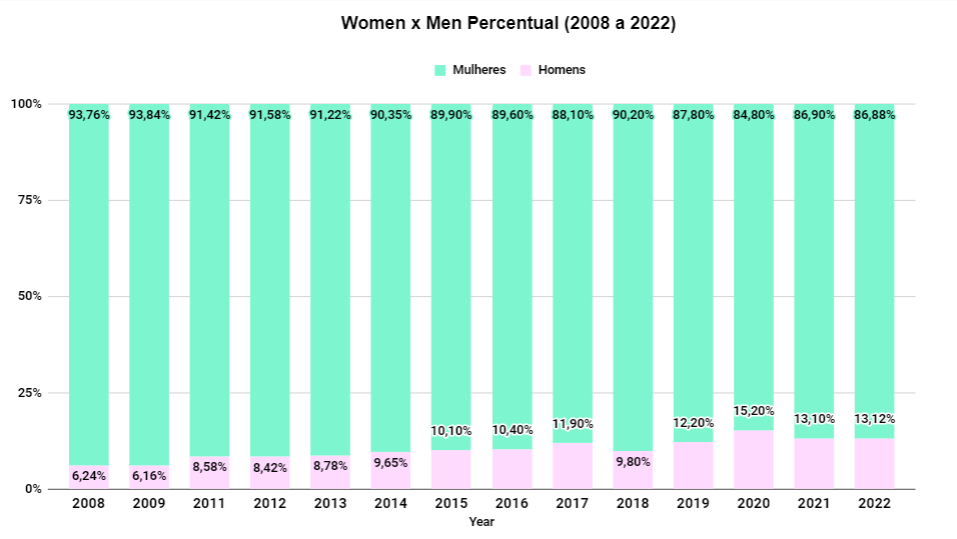}
\caption{Comparison of Men and Women (2008 a 2022)}
\label{fig:exampleFig4}
\end{figure}

This scenario, which is not exactly hopeful where women have not even 15\% marathons participation, is even worse when there is a real comparison in relation to male percentage (Figure~\ref{fig:exampleFig4}).

\begin{figure}[ht!]
\centering
\includegraphics[width=.95\textwidth]{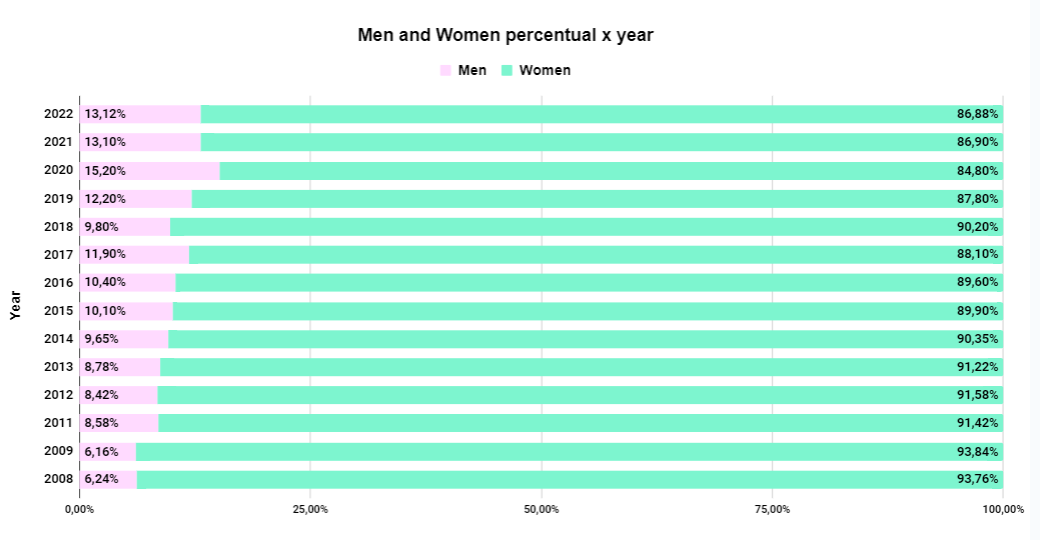}
\caption{Comparison of Women versus Men by year}
\label{fig:exampleFig5}
\end{figure}

\cite{Cursino:21}, in their article “Descriptive Statistical Analysis and Regression of the Insertion of Women in IT Courses in the Years 2009 to 2018”, analyzing the female profile in IT courses from 2009 to 2018, there is a strong positive correlation in relation to the male gender, showing that those in a positive and proportional growth, but at the same time, negative in relation to the female gender, inversely correlated, move in opposite directions, but proportional, (Figure~\ref{fig:exampleFig5}). 
The same occurs when analyzing the global perception of Programming Marathons. The number of participants has grown over the years, but according to (Figure~\ref{fig:exampleFig5}) it can be seen that inequality between women and men remains.

From a general perspective, at times, such as in 2020, there was an increase in female participants, which represents an increase of 100\% compared to 2008. It was also the most significant year in terms of female participation. However, the optimism that one has in relation to this growth did not, in fact, have a really significant representation in the general context, because as the colloquial say, “twice as much is still nothing”. Female participation is derisory in relation to male participation as a whole. It has no proportionality, even in most years, with female participation in computing courses ranging from 13.8\% to a maximum of 15.2\%, as mentioned in the study by \cite{Santos:21}.

\begin{figure}[ht!]
\centering
\includegraphics[width=.8\textwidth]{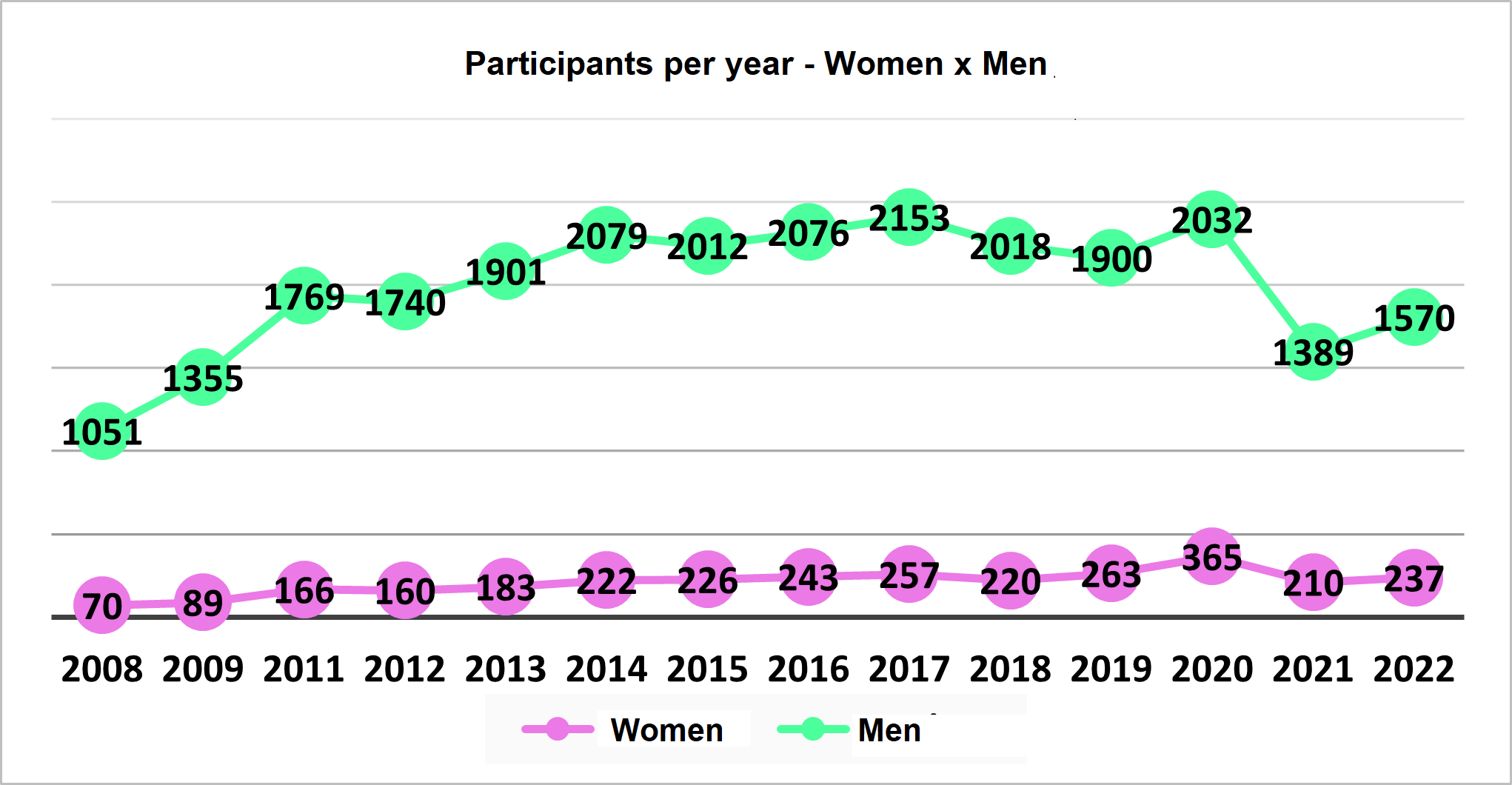}
\caption{Women x Men Comparison}
\label{fig:exampleFig6}
\end{figure}

Still observing the data, in a comparative analysis in a line graph of women participating in the Programming Marathons in relation to men, one can see the general growth over the last few years, but  women remain with a minimum of representations, like to seeing at (Figure~\ref{fig:exampleFig6}) \cite{SBC:23}.
 
In view of what was exposed in the results of recent years in the Programming Marathons, the need for incentive actions for the inclusion of women in these types of competitions is evident.

\section{Conclusions and future work}
Female representation in Programming Marathons is a recurring discussion among participants in these events. In general, women are a minority in competitions, which raises the issue of inclusion and equal opportunities.

Some women have stood out in the Programming Marathons, winning important victories. However, there is still a long way to go for female representation to be truly egalitarian in this environment.

The lack of representation and empowerment of women in the technology industry has been a major concern, but the participation of girls is a positive step towards a more balanced and diverse industry. In addition, it is important to highlight that the girls participation in Programming Marathons is crucial to change the social perception of the role of women in technology and to show that they have potential to be leaders and innovators in this area.

Women's participation in Programming Marathons can also boost their self-esteem and confidence in the field of technology. By competing and gaining recognition, they can be inspired to pursue careers in technology and make a difference to the world.

A factor that may contribute to the increase in the number of women in the field of technology may be precisely the greater involvement of other women and their prominence in these events. Inspiring women show that it is possible to be a successful woman in technology field.

In summary, Programming Marathons are an important vehicle for the representation and empowerment of girls in the tech industry. Increasing women's participation in these events is crucial to achieving diversity and gender equality in the industry and to changing societal perceptions of the role of women in technology.

As future work, the proposal for the study and development a method that can encourage and motivate (and keep motivated) that seeks to bring women closer to intellectual programming competitions, in particular the Programming Marathons.

\bibliographystyle{sbc}
\bibliography{wit}

\end{document}